\documentclass[aip, amsmath,amssymb, reprint]{revtex4-1}

\usepackage{graphicx}
\usepackage{dcolumn}
\usepackage{bm}

\usepackage[utf8]{inputenc}
\usepackage[T1]{fontenc}
\usepackage{mathptmx}
\usepackage{etoolbox}
\usepackage{color}

\begin{document}

\preprint{AIP/123-QED}

\title[Giant demagnetization effects induced by superconducting films]{Giant demagnetization effects induced by superconducting films}
\author{S. V. Mironov}
\affiliation{Institute for Physics of Microstructures, Russian Academy of Sciences, 603950 Nizhny Novgorod, GSP-105, Russia}
\author{A. I. Buzdin}
\email[Author to whom correspondence should be addressed: ]{alexandre.bouzdine@u-bordeaux.fr}
\affiliation{University Bordeaux, LOMA UMR-CNRS 5798, F-33405 Talence Cedex, France}
\affiliation{World-Class Research Center ``Digital biodesign and personalized healthcare'', Sechenov First Moscow State Medical University, Moscow 119991, Russia}

\date{\today}

\begin{abstract}
We show that a ferromagnetic (F) slab with the in-plane magnetization sandwiched between two superconducting (S) films experiences strong demagnetization effect due to the Meissner screening of the stray magnetic field by the superconductors. In the extreme case the transition of the S films from normal to the superconducting state can switch the demagnetization factor from $0$ to $1$ which is in a sharp contrast with the S/F bilayers where such transition affects the magnetic field inside the F film only slightly. The giant demagnetization effect is shown to be qualitatively robust against the decreasing of the superconducting film thickness and may provide a hint towards the explanation of the anomalously large ferromagnetic resonance frequency shift recently observed for the S/F/S structures [I. A. Golovchanskiy, N. N. Abramov, V. S. Stolyarov, V. I. Chichkov, M. Silaev, I. V. Shchetinin, A. A. Golubov, V. V. Ryazanov, A. V. Ustinov, and M. Yu. Kuprianov,  Phys. Rev. Appl. \textbf{14}, 024086 (2020)].
\end{abstract}

\maketitle

In recent years the incoming experimental data on the electrodynamics of superconductor (S) / ferromagnet (F) hybrids have uncovered several puzzling phenomena contradicting the common beliefs \cite{Flokstra_1, Flokstra_2, Flokstra_3, Flokstra_4, Khaydukov, Golovchanskiy_1, Li, Blamire_1, Blamire_2}. The basic mechanism of electromagnetic interaction between these two kinds of materials is associated with the Meissner effect responsible for the expulsion of the magnetic field from the bulk of superconductors \cite{Tinkham}. Specifically, the stray magnetic field induced by domain structure or edges of ferromagnet penetrates the adjacent superconductor and generates the screening Meissner currents there. These currents can have a back-action to the ferromagnet and the resulting modifications of the magnetic patterns in the F sample have been the subject of extensive theoretical and experimental analysis \cite{Aladyshkin}. It is natural to assume that the described interaction should become damped for the large F films with in-plane uniform magnetization since their stray magnetic fields are localized near the edges and are negligibly small at the central region of the film. However, the recent experiments showed that it is not the case. Specifically, muon spin rotation techniques and neutron reflectometry measurements performed for the multilayered S/F structures with in-plane magnetization and good electric contact between the layers detect the penetration of magnetic field from the F to the S layer over the distances strongly exceeding the diffusion lengths for the spin-polarized electrons \cite{Flokstra_1, Flokstra_2, Flokstra_3, Flokstra_4, Khaydukov}. Also, the ferromagnetic resonance (FMR) measurements show the giant shifts in the FMR frequency for the S/F/S trilayers below the critical temperature of the superconducting phase transition while for the S/F bilayers the FMR frequency shifts appear to be negligible \cite{Golovchanskiy_1, Li, Blamire_1, Blamire_2}. 

This unexpectedly strong influence of superconductors on magnetic fields induced by ferromagnets presents the challenge for the existing theory. The possible explanation for the long-range magnetism in S/F structures with electrically transparent interfaces is based on the theory of electromagnetic proximity effect \cite{Mironov, Devizorova, Volkov}. The Cooper pairs penetrating the F layer due to the proximity effect screen the magnetization induced magnetic field \textit{inside} the ferromagnet. The corresponding Meissner currents in the F layer become compensated by the supercurrent flowing inside the superconductor, which gives rise to the magnetic field decaying over the London penetration depth inside the S layer. Although this theory provides a comprehensive explanation for the long-range electromagnetic effect, it struggles to explain the giant FMR frequency shifts in S/F/S structures. 

\begin{figure}[b!]
\begin{center}
\includegraphics[width=0.7\linewidth]{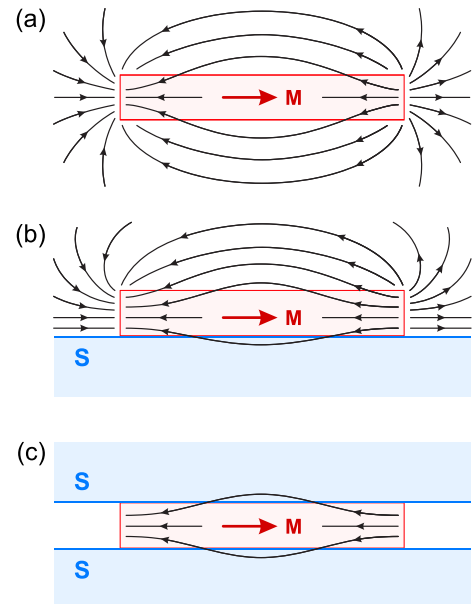}
\end{center}
\caption{Sketch of the magnetic field intensity ${\bf H}$ profiles for (a) isolated ferromagnetic slab, (b) ferromagnetic slab on top of superconductor, (c) ferromagnetic slab sandwiched between two superconductors. In all panels the thickness of the slab well exceeds $\lambda$.}\label{Fig_System}
\end{figure}

The aim of this Letter is to demonstrate some unusual behavior of the magnetic field distribution in ferromagnets sandwiched between two superconducting films (Fig.~\ref{Fig_System}). The presence of the second superconducting film qualitatively modify the demagnetization effect and may even switch the demagnetization factor from $0$ to $1$. We believe that these important changes of the internal fields may provide the hints for the explanation of the recent experimental results concerning the anomalous FMR frequency shift in S/F/S trilayers \cite{Golovchanskiy_1, Li, Blamire_1, Blamire_2}.

\begin{figure}[t!]
\begin{center}
\includegraphics[width=0.75\linewidth]{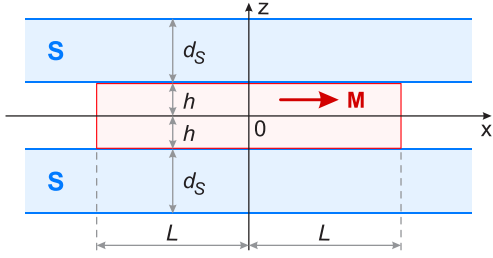}
\end{center}
\caption{Geometry of the system consisting of the ferromagnetic slab sandwiched between two superconducting films.}\label{Fig2}
\end{figure}

To elucidate our key idea let us consider the infinite ferromagnetic slab of the rectangular cross-section $2h\times 2L$ placed (a) in a vacuum, (b) on top of the infinite S film, and (c) between two S films. For all three cases we will use the coordinate system shown in Fig.~\ref{Fig2} [the sketch corresponds to the case (c) while for two other cases one should omit upper or both S layers]. The magnetization ${\bf M}$ inside the slab is assumed to be uniform and directed along the $x$ axis. 

Let us start from the case of an isolated ferromagnet. If the sample width $L\to\infty$ the magnetic field induction is trapped inside the ferromagnet ${\bf B}=4\pi{\bf M}$ while outside the slab ${\bf B}=0$. At the same time, the slabs with finite $L$ reveal demagnetization effects associated with the stray magnetic field (Fig.~\ref{Fig_System}a). In what follows we will be focused on the case of the film geometry considering the slabs with $L\gg h$. To calculate the magnetic field inside and outside the sample it is convenient to introduce the magnetic intensity ${\bf H}={\bf B}-4\pi{\bf M}$ satisfying the magnetostatics equations $\nabla\times{\bf H}=0$ and $\nabla\cdot{\bf H}=4\pi\rho$ where $\rho=-\nabla\cdot{\bf M}$ is the magnetic monopole charge density. For the F slab the magnetic monopoles are uniformly distributed along the boundaries $x=\pm L$ with the surface density $\sigma=\pm M$, respectively, and the induced stray field ${\bf H}$ lays in the $xz$ plane. At distances $r\gg h$ from the ``charged'' edges the stray field ${\bf H}$ can be well approximated by the field induced by the two parallel wires charged with the density $\beta=\pm 2Mh$ and located at $x=\pm L$ and $z=0$. The field from each wire is directed in radial direction and decays as $H=2\beta/r$. As a result, if $L\gg h$ the total magnetic field ${\bf B}$ in the central part of the slab (e.g. at $x=z=0$) is weakly affected by the stray magnetic field and the demagnetization effects are observable only at the distance $\sim h$ near the edges $x=\pm L$. 

The very similar situation is realized when the F slab is put on top of the infinite S film of the thickness well exceeding the London penetration depth $\lambda$ (see Fig.~\ref{Fig_System}b). We will be interested only in the electromagnetic interaction between the layers, so for simplicity we assume that the S/F interface is insulating and we can neglect a superconducting proximity effect. Also we assume that $2h\gg\lambda$ which ensures that only small part of the magnetic field lines penetrate to the superconductor. The stray magnetic field induced by the edges of  F slab tends to penetrate the superconductor. However, inside the S film this field becomes fully screened by the Meissner currents flowing in the surface region of the thickness $\sim \lambda$, so that in the bulk of the superconductor ${\bf B}=0$. In turn, the Meissner currents also produce the magnetic field ${\bf H}_s$ in the outer space which doubles the diamagnetization effect in the F slab. To show this let us again consider the F slab and the lower superconductor (located at the region $z<-h$) shown in Fig.~\ref{Fig2}. To calculate the field ${\bf H}_s$ one can apply some sort of image method: each real magnetic charge $q_m$ located at the point $(x,y,z)$ above the edge of the superconductor (i.e. for $z>-h$) should be accompanied by the image charge $q_m$ of the same sign located at the point $(x,y,-2h-z)$, and the field ${\bf H}_s$ in the region $z>-h$ coincides with the field produced by all image charges. As a result, for the ferromagnetic slabs with $L\gg h$ the account of image charges when calculating the stray field ${\bf H}$ effectively doubles the linear charge density $\beta$ of the magnetic wires located at $x=\pm L$ which leads to the doubling of the field component  $H_x$ in the central part of the F film compared to the case of the isolated ferromagnet. Of course, in this case the demagnetization effects are, again, negligibly small far from the edges $x=\pm L$.

Remarkably, the demagnetization effects becomes dramatically enhanced provided the ferromagnetic slab of the thickness $2h\gg\lambda$ is sandwiched between two superconductors (see Fig.~\ref{Fig_System}c). In this case, the magnetic charges at the ferromagnet edges $x=\pm L$ (see Fig.~\ref{Fig2}) induce the infinite series of the image charges in the regions of superconductors. Consequently, the field ${\bf H}$ induced inside the ferromagnet now coincides with the field from two parallel magnetic planes (``magnetic capacitor'') perpendicular to the $x$ axis and located at $x=\pm L$. This field is equal to ${\bf H}=-4\pi {\bf M}$ and does not decay from the slab edges which is in sharp contrast to the above case of only one superconducting film. As a result, one gets perfect demagnetization inside the slab geometry ferromagnet, so that ${\bf B}=0$. 

Note that accounting the finite size $2L_y$ of the F slab in the $y$ direction does not change the results qualitatively provided $L_y\gg L$. In this case the field ${\bf B}$ in the central part of the ferromagnet (i. e. at distances much larger than $h$ from the slab edges $y=\pm L_y$) should be of the order of $B\sim M (L/L_y)\ll M$ (specifically, at the coordinate system origin $B_x/M\approx 8L/L_y$).  Also the above effect should be qualitatively robust towards the decreasing of the F slab thickness $2h$. In the case $2h\sim\lambda$ one may expect that almost half of the magnetic field lines penetrates the S layers and the demagnetization factor drops from 1 down to $\sim 1/2$.

The giant demagnetization effects reveal themselves even if the thickness of the S films $d_s$ covering the F slab is comparable or smaller than the London penetration depth $\lambda$. To demonstrate this we consider the limiting case when $d_s\ll\lambda$  (Pearl's limit \cite{Pearl}) and calculate the stray magnetic field ${\bf H}$ using the approach from Ref.~\cite{Vlasko-Vlasov}. We choose the vector potential ${\bf A}=A(x,z)\hat{\bf y}$ corresponding to the magnetic field ${\bf B}=\nabla\times{\bf A}$ can be considered as a sum of two contributions: ${\bf A}={\bf A}_m+{\bf A}_s$. The first component ${\bf A}_m$ is induced by the ferromagnet while the second one ${\bf A}_s$ is induced by the superconducting currents ${\bf j}_s$ flowing inside the S films. According to the Maxwell equations, $\nabla\times\left(\nabla\times{\bf A}_s\right)=(4\pi/c){\bf j}_s$ where the current ${\bf j}_s$ is defined by the total local vector potential through the London relation \begin{equation}\label{London}
{\bf j}_s=-(c/4\pi\lambda^2){\bf A}.
\end{equation}
Note that since the magnetization profile ${\bf M}({\bf r})$ inside the ferromagnet satisfies the symmetry relation $M_x(x,-z)=M_x(x,z)$  the vector potential $A({\bf r})$ can be chosen in a way that $A(x,-z)=-A(x,z)$.

In the limit $d_s\ll\lambda$ the superconducting films can be considered as delta-layers so that we obtain the following equation for the vector potential ${\bf A}_s$:
\begin{equation}\label{Maxwell}
\left(\partial^2_x+\partial^2_z\right) A_{s}=\frac{d_s}{\lambda^2}\left[\delta(z-h)+\delta(z+h)\right]A.
\end{equation}
To solve Eq.~(\ref{Maxwell}) is it convenient to perform the Fourier transform for all components $a(x,z)$ of the vector potential ($a$ can be either $A$, $A_s$ or $A_m$) and also for the profiles $a(x,h)$ inside the upper superconducting film:
\begin{equation}\label{Fourier_def}
a({\bf k})=\int a({\bf r})e^{i{\bf k}{\bf r}}d^2{\bf r},~\bar{a}(k_x)=\int a(x,h)e^{ik_xx}dx.
\end{equation}
Here ${\bf r}=(x,z)$ and ${\bf k}=(k_x,k_z)$ are the two-dimensional vectors in the $xz$ plane. Then Eq.~(\ref{Maxwell}) takes the form
\begin{equation}\label{StartEq}
 A_{s}({\bf k})=-\frac{2i\sin(k_zh)}{\lambda_{\rm eff}\left(k_x^2+k_z^2\right)}\bar{A}(k_x),
\end{equation}
where $\lambda_{\rm eff}=\lambda^2/d_s$ is the effective Pearl length. Performing the inverse Fourier transform with respect to only $z$ coordinate in Eq.~(\ref{StartEq}) one may write down the expression for the vector potential component $A_s(k_x,z)$. Substituting $z=h$ into the obtained expression and taking into account that $\bar{A}_s(k_x)=\bar{A}(k_x)-\bar{A}_m(k_x)$ we obtain the algebraic equation for $\bar{A}(k_x)$: 
\begin{equation}\label{APM}
\bar{A}(k_x)-\bar{A}_m(k_x)=-\frac{\bar{A}(k_x)}{2\pi\lambda_{\rm eff}}\int \frac{1-e^{-2ik_zh}}{k_x^2+k_z^2}dk_z.
\end{equation}
Then taking the integral in Eq.~(\ref{APM}) we find
\begin{equation}\label{APMsol}
\bar{A}(k_x)=\bar{A}_m(k_x)\left(1+\frac{1-e^{-2\left|k_x\right|h}}{2\left|k_x\right|\lambda_{\rm eff}}\right)^{-1}
\end{equation}
and from Eq.~(\ref{StartEq}) obtain $A_s({\bf k})$.  

The $x$ component of the magnetic field ${\bf B}_s={\rm rot}{\bf A}_s$ induced by the supercurrents in the plane $z=0$ reads
\begin{equation}\label{B_def}
B_{sx}(x,0)=(2\pi)^{-2}\int ik_zA_{s}({\bf k})e^{-ik_xx}d^2{\bf k}.
\end{equation}
Taking the integral with respect to $k_z$ we find:
\begin{equation}\label{Bs_res1}
B_{sx}(x,0)=(2\pi\lambda_{\rm eff})^{-1}\int \bar{A}(k_x)e^{-\left|k_x\right|h-ik_xx}dk_x.
\end{equation}

In the limit $L\gg h$ one can model the field from the F slab by the field induced by two magnetically charged wires. The vector potential of a wire with the linear charge density $\beta$ reads $A_w=2\beta\varphi$, where $\varphi$ is the azimuthal angle around the wire (the increase of $\varphi$ corresponds to the clockwise rotation in Fig.~\ref{Fig2}). So, the vector potential induced by the two wires with the opposite charges positioned at $x=\pm L$ and $z=0$ inside the upper superconducting film (at $z=h$) takes the form 
\begin{equation}\label{A_wire}
A_m(x,h)=-4Mh\left[{\rm arctg}\left(\frac{x+L}{h}\right)-{\rm arctg}\left(\frac{x-L}{h}\right)\right],
\end{equation}
while for the lower S film the expression for $A_m$ should account the finite magnetic field inside the F slab: $A_m(x,h)-A_m(x,-h)=-\int_{-h}^hB_x(x,z)dz$. The corresponding stray magnetic field decays at distances $\sim h$ from each wire so that in the central region of the ferromagnet (for $\left|x\right|\ll L$) it is small: $H_{mx}\approx -8Mh/L$. However, the magnetic field ${\bf B}_s$ induced by the superconducting currents is not negligibly small even near $x=0$. To demonstrate this it is enough to consider the limit $k_x\to 0$ in Eq.~(\ref{Bs_res1}) since all other Fourier components (with non-zero $k_x$) should average down to zero far from the charged wires. In this limit accounting that $h\ll \lambda_{\rm eff}$ from Eq.~(\ref{APMsol}) one finds
\begin{equation}\label{APMsol2}
\bar{A}(k_x\to 0)\approx \bar{A}_m(k_x)\left(1+h/\lambda_{\rm eff}\right)^{-1}\approx \bar{A}_m(k_x)
\end{equation}
and, thus, for $\left|x\right|\ll L$
\begin{equation}\label{Bs_res2}
B_{sx}(x,0)\approx\int \frac{\bar{A}_m(k_x)e^{-ik_xx}dk_x}{2\pi\lambda_{\rm eff}} =\frac{A_m(x,h)}{\lambda_{\rm eff}}\approx -\frac{4\pi Mh}{\lambda_{\rm eff}}.
\end{equation}
Remarkably, this magnetic field does not depend on $L$ and for large F slabs with $L\gg \lambda_{\rm eff}$ it substantially exceeds the stray magnetic field $H_{mx}$ induced by the ferromagnet. 

The above results are valid for the ferromagnetic slab sandwiched between \textit{infinite} superconducting films while in typical S/F heterostructures the superconducting layer covering the ferromagnet has the same lateral size. However, the approximate solution of the Maxwell equations for the half-infinite (positioned at $x>0$) ferromagnetic slab of the thickness $2h$ sandwiched between two half-infinite superconducting films of the thickness $d_s\ll \lambda$ shows that far from the edge of the sample $x=0$ the $x$-component of the magnetic field ${\bf B}_s$ in the plane $z=0$ reads $B_{sx}\approx -2\pi M(h/\lambda_{\rm eff})$ (see supplementary material). This result is only twice smaller than the one for the case of the infinite superconducting films. Note that the same expression for the demagnetization field is valid for the ferromagnetic slab sandwiched between two superconducting films of the same width $2L$: in the middle F layer the demagnetization field $B_{sx}\approx -2\pi M(h/\lambda_{\rm eff})$, which may strongly exceed an usual demagnetization field $\sim -M\left(h/L\right)$, provided $L>>\lambda_{\mathrm{eff}}$. We see that the presence of superconducting film introduces an additional length $\ \lambda _{\mathrm{eff}}$: the demagnetization effect is strongly enhanced when $L\gg\lambda_{\mathrm{eff}}$. Note that this conditions if fulfilled in the experiments \cite{Golovchanskiy_1, Li, Blamire_1, Blamire_2}. Note also that in the considered situation  the total current in S layer is not zero. It means that the current loops should close from upper to lower films somewhere at infinity ($y\rightarrow \pm \infty $), in other words it will depend on the overall geometry of the system. If we impose the requirement of the vanishing of the  total current the additional demagnetization field drops to (see supplementary material)
\begin{equation}
B_{sx}\sim -M\left(\frac{h}{\lambda _{\mathrm{eff}}}\right)\left(\frac{h}{L}\right)\ln \left(\frac{2L}{h}\right).
\end{equation}

In practice for the small size sample the current loops could close through a Josephson current at the borders of the sample. We note also that the FMR experiments \cite{Golovchanskiy_1, Li, Blamire_1, Blamire_2} were performed at the range of
GHz frequencies at we may suppose that the presence of the capacitive coupling between the layers would provide an effective way to shunt them at these frequencies.

Theoretical analysis by Kittel \cite{Kittel} provides the following formula for the FMR frequency (for convenience, we renormalize the demagnetization factors):
\begin{equation}\label{FMR}
\omega_{0}=\gamma \sqrt{\left[H_{x}+4\pi (N_z-N_{x})M_{x}\right] \left[
H_{x}+4\pi\left( N_{y}-N_{x}\right) M_{x}\right] },
\end{equation}
where $\gamma$ is the gyromagnetic ratio. For the sample of the plate-like form with the $z$ axis perpendicular to the plane and the magnetic field parallel to the magnetic moment, oriented along the $x$ axis one may put $N_z=1$. In the normal state the demagnetization factors $N_{y}\approx N_{x}\approx 0$, while the transition of the S layers into the superconducting state should generate a specific demagnetizing field as previously discussed and then provide a shift of the resonance field. This shift is determined by the interplay between the demagnetization effect
induced by superconductivity along $x$ and $y$ direction which should depend on the details of the reconnection of the screening current loops. The demagnetization effect along the $x$ axis is related with a static field $M_x$ and in the case of the vanishing total static current along the $y$ direction it is quite small, so $N_x$ remain small (almost zero). On the contrary, the demagnetization effect along the $y$ axis is related with a dynamic GHz field $M_y$ and due to capacitive coupling between the layers the demagnetization factor $N_y$ may be close to 1. This circumstance should lead to the increase of the resonance frequency in the superconducting state in accordance with the experimental observations [\onlinecite{Golovchanskiy_1, Li, Blamire_1, Blamire_2}].

Thought it is rather difficult to model the screening current
distribution for the experimental setups \cite{Golovchanskiy_1, Li, Blamire_1, Blamire_2}, we may expect that the discussed demagnetization effect should substantially increase when the size of S layers will exceed that of F layer. It could be interesting to check this prediction on experiment.

\begin{figure}[t!]
\begin{center}
\includegraphics[width=0.6\linewidth]{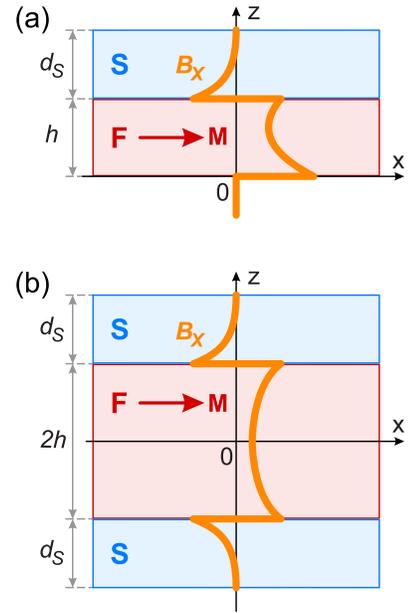}
\end{center}
\caption{Sketch of the magnetic field profiles $B_x(z)$ induced in (a) S/F and (b) S/F/S structures due to the electromagnetic proximity effect.}\label{Fig3}
\end{figure}

In the above theory we neglected the proximity effect coming from the possible electron transfer between F and S layers. Interestingly, such proximity effect provides an additional mechanism of the strong demagnetization effect. To demonstrate this let us compare the magnetic field inside the ferromagnet for the planar S/F and S/F/S structures shown in Fig.~\ref{Fig3}. For simplicity we consider the F layer thickness $d_f\lesssim \xi_f$ and neglect the variation of the Cooper pair wave function inside the ferromagnet (here $\xi_f$ is the superconducting coherence length inside the ferromagnet). Then the London penetration depth $\lambda_f$ in the F layer is constant but, in principle, it can differ from the one $\lambda_s$ inside the superconductor. The penetration of Cooper pairs to the F layer is known to induce the so-called electromagnetic proximity effect \cite{Mironov}. In the absence of superconductors the ferromagnetic film of the infinite lateral size and in-plane magnetization ${\bf M}$ does not generate the stray magnetic field in the outer space while inside the F film the magnetic induction ${\bf B}_0=4\pi {\bf M}$. However, in S/F or S/F/S structures with the proximity effect the field ${\bf B}_0$ becomes screened by the Meissner currents flowing {\it inside} the ferromagnet. This effect is accompanied by the generation of the counterflowing currents in the adjacent superconductors and the emergence of the magnetic field penetrating the S layers over the distances $\sim\lambda_s$. In what follows we demonstrate that the damping of the magnetic induction ${\bf B}$ inside the ferromagnet due to the electromagnetic proximity effect is much stronger for the S/F/S structures compared to S/F bilayers.

The geometry of the heterostructures is sketched in Fig.~\ref{Fig3}. The magnetization ${\bf M}$ inside the F layer is directed along the $x$ axis, so the magnetic field ${\bf B}$ in the whole structure has only one component $B_x$ which depends on the coordinate $z$. To find the profile $B_x(z)$ we solve the London equation $\nabla\times\left[\lambda^2(z)\nabla\times{\bf B}\right]+{\bf B}=0$ where the London penetration depth takes the values $\lambda_s$ and $\lambda_f$ in the S and F layers, respectively. The field ${\bf B}$ is continuous at the interfaces between the S layers and vacuum while at the edges of the F layer it experiences the jumps $\pm 4\pi M$. Also at the S/F interfaces one needs to ensure the continuity of the combination $\lambda^2 (\partial B_x/\partial z)$ (which follows from the vector potential continuity). In the absence of external magnetic field applied to the system the solution of the London equation for the S/F bilayer takes the form
\begin{equation}\label{EPE_B_SF}
B_x=\left\{ 
\begin{array}{l}{\displaystyle 4\pi M \cosh\left(z/\lambda_f\right)+W_1\sinh\left(z/\lambda_f\right), ~0<z<h,}
\\{\displaystyle W_2\sinh\left[\left(z-h-d_s\right)/\lambda_s\right], ~h<z<h+d_s,}\end{array}
\right.
\end{equation}
where the constants $W_1$ and $W_2$ are determined by the boundary conditions at the S/F interface (see supplementary material for the explicit expressions). In the limiting case $h\ll\lambda_f$ and $d_s\ll\lambda_s$ which is typical for most experiments the demagnetization field at the S/F interface is $\Delta B_{SF}=B_x(h)-4\pi M\approx -4\pi M (h/\lambda_f)$. At the same time, for the S/F/S trilayers with the F layer thickness $2h$ the field profile for $z>0$ reads
\begin{equation}\label{EPE_B_SFS}
B_x=\left\{ 
\begin{array}{l}{\displaystyle W_3  \cosh\left(z/\lambda_f\right), ~0<z<h,}
\\{\displaystyle W_4\sinh\left[\left(z-h-d_s\right)/\lambda_s\right], ~h<z<h+d_s,}\end{array}
\right.
\end{equation}
and $B_x(-z)=B_x(z)$ (the expressions for $W_3$ and $W_4$ are presented in supplementary material). For such structures in the limit $h\ll\lambda_f$ and $d_s\ll\lambda_s$ one obtains that at the S/F interface the demagnetization field $\Delta B_{SFS}\approx -4\pi M (h/\lambda_s)$. The absence of the superconducting paring potential for the electrons inside the ferromagnet and possible barriers for the electron transfer at the S/F interfaces should result in the low superfluid density inside the F layer as compared to the superconducting films and, thus, it is natural to assume $\lambda_f\gg\lambda_s$. Then one obtains $\Delta B_{SFS}/\Delta B_{SF}\sim \lambda_f/\lambda_s\gg 1$ which confirms that the demagnetization effects in S/F/S structures is much stronger than the ones in S/F bilayers. Note that the obtained estimate for the demagnetization field in the limit $h\ll\lambda_f$ coincides with Eq.~\ref{Bs_res2} where $\lambda_{\rm eff}$ is replaced with $\lambda_s$. Both mechanisms, namely the screening of the stray magnetic fields inside the S layers and the electromagnetic proximity effect are based on the orbital interaction between the magnetic field and the superconducting condensate. However, these two mechanisms reveal themselves completely different in the case when $h\gtrsim \xi_f$ ($\xi_f$ is the superconducting coherence length inside the ferromagnet): the contribution from the electromagnetic proximity effect becomes damped with the increase in $h$ while the effect coming from the screening of the stray fields is not sensitive to the F thickness growth which makes it dominant.  In \cite{Golovchanskiy_1} the effect was observed even in the samples with a thickness of F layer up to 350 nm. This thickness strongly exceeds the distance of the penetration of superconducting correlations in ferromagnetic, which is of few nm, and then the S layers should be completely decoupled. Then this electromagnetic proximity effect cannot explain the results of the experiments \cite{Golovchanskiy_1, Li, Blamire_1, Blamire_2}.


Therefore, we believe that the origin of the interplay between superconductivity and FMR observed in Refs.~\onlinecite{Golovchanskiy_1, Li, Blamire_1, Blamire_2} is related to the giant demagnetization effect produced by two S films. Following our results [see Eq.~(\ref{Bs_res2})] the modification of the internal field in a ferrimagnet increases with the increase of its thickness $h$ which is in accordance with the results of Ref.~\onlinecite{Golovchanskiy_1}. In the limit $d_s<\lambda_s$ the shift of the FMR frequency should be proportional to $d_s$ and it may be interesting to check this prediction on experiment. More generally, the revealed giant demagnetization effect should be taken into account in the design of devices for superconducting spintronics.

\section*{Supplementary Material}

Detailed calculations of the demagnetization field in S/F/S sandwiches with superconductors of finite lateral size and detailed analysis of the electromagnetic proximity effect in S/F and S/F/S systems.

\begin{acknowledgments}
The authors thank V. V. Ryazanov, A. S. Mel'nikov, A. V. Samokhvalov and A. Yu. Aladyshkin for useful discussions. This work was supported by the French ANR OPTOFLUXONICS and EU COST CA16218 Nanocohybri. A.I.B. acknowledges support by the Ministry of Science and Higher Education of the Russian Federation within the framework of state funding for the creation and development of World-Class Research Center “Digital biodesign and personalized healthcare” N075-15-2020-92.
\end{acknowledgments}

This article may be downloaded for personal use only. Any other use requires prior permission of the author and AIP Publishing. This article appeared in S.~V.~Mironov and A.~I.~Buzdin, Appl. Phys. Lett. 119, 102601 (2021) and may be found at https://doi.org/10.1063/5.0059149.

\renewcommand{\theequation}{S\arabic{equation}}

\section*{Supplementary material for ``Giant demagnetization effects induced by superconducting films''}

\setcounter{equation}{0}

\subsection{Demagnetization effect of S/F/S sandwiches with superconductors of finite lateral size}

To analyze how the finite size of the superconducting films affects the long-range demagnetization phenomenon let us consider the half-infinite (in the $x$ direction) ferromagnetic slab of the thickness $2h$ sandwiched between two half-infinite superconducting films of the thickness $d_s\ll \lambda$. Let us assume that the F and S films are positioned in the region $x>0$ so that the plane $x=0$ corresponds to the edge of the sample. For simplicity the size of the slab in the $y$ direction is assumed to be infinite. The magnetization ${\bf M}$ is assumed to be directed along the $x$ axis so that $\nabla\cdot{\bf M}\neq 0$ only at the slab edge at $x=0$. 

The superconducting currents ${\bf j}_s$ screening the stray magnetic field flow along the $y$ axis and due to the system symmetry ${\bf j}_s(x,-h)=-{\bf j}_s(x,h)$. The relation between the current density ${\bf j}_s$ and the total vector potential for $x>0$ is defined by the Pearl relation 
\begin{equation}\label{CurrPearl}
{\bf j}_s(x,\pm h)=-\frac{c}{4\pi\lambda_{\rm eff}}\delta(z\mp h){\bf A}(x,z). 
\end{equation}
To calculate the part of the vector potential ${\bf A}_s$ induced by these superconducting currents we use the Bio-Savar-Laplas expression
\begin{equation}\label{BSL}
{\bf A}_s({\bf r})=\frac{1}{c}\int \frac{{\bf j}_s(x^\prime,z^\prime) dx^\prime dy^\prime dz^\prime}{\sqrt{(x-x^\prime)^2+y^{\prime 2}+(z-z^\prime)^2}}, 
\end{equation}
where the integral is taken over the half-space $x>0$. Substituting Eq.~(\ref{CurrPearl}) into Eq.~(\ref{BSL}), taking the integral over $y^\prime$ and accounting that ${\bf j}_s(x,-h)=-{\bf j}_s(x,h)$ we find:
\begin{equation}\label{IEG}
\begin{array}{c}{\displaystyle
A_s({\bf r})=-\int \frac{A(x^\prime,h)}{4\pi\lambda_{\rm eff}} \left(\frac{1 }{\sqrt{(x-x^\prime)^2+y^{\prime 2}+(z-h)^2}}\right.}\\{\displaystyle \left.-\frac{1 }{\sqrt{(x-x^\prime)^2+y^{\prime 2}+(z+h)^2}}\right)dx^\prime dy^\prime}.\end{array} 
\end{equation}
Now the integral over $y^\prime$ can be calculated explicitly. Note that the r.h.s. of Eq.~(\ref{IEG}) is proportional to the small parameter $h/\lambda_{\rm eff}\ll 1$ which allows to find $A_s({\bf r})$ perturbatively. In the first order over $h/\lambda_{\rm eff}$ we may put $A(x^\prime,h)\approx A_m(x^\prime,h)$ under the integral. Then
\begin{equation}\label{IEAs}
A_s(x,z)=\int\limits_0^\infty \frac{A_m(x^\prime,h)}{4\pi\lambda_{\rm eff}} {\rm ln}\left[\frac{(x-x^\prime)^2+(z-h)^2}{(x-x^\prime)^2+(z+h)^2}\right]dx^\prime. 
\end{equation}
The $x$-component of the magnetic field ${\bf B}_s=\nabla\times{\bf A}_s$ in the plane $z=0$ is equal to $B_{sx}=-\partial A_s/\partial z$. Taking $A_m(x^\prime,h)=-4Mh~{\rm arctg}\left(x^\prime/h\right)$ we find that for $x\gg h$
\begin{equation}\label{Bs}
B_{sx}(x,0)=-\frac{4Mh}{\pi\lambda_{\rm eff}}\int\limits_0^\infty \frac{{\rm arctg}\left(t\right)dt}{1+(t-x/h)^2}\approx -\frac{2\pi Mh}{\lambda_{\rm eff}}.
\end{equation}
This result is twice smaller than the one for the case of the infinite superconducting films.

Similar perturbative approach gives the vector potential $A_{s}(x,z)$ generated by the ferromagnetic slab sandwiched between two superconducting films of the same width $2L$ - we should just use the formula (\ref{IEAs}) with $A_{m}(x,h)$ given by (9) and perform the integration over $x^{\prime }$ in the interval $-L<x^{\prime}<L$.

Note that in the considered situation the total current in S layer is not zero. It means that the current loops should close from upper to lower films somewhere at infinity ($y\to\pm\infty$).  If we impose the requirement of the vanishing of the total current the result for the additional demagnetization field will be different. Let us calculate the field ${\bf B}_s$ for the F slab of the cross-section $2L\times 2h$ sandwiched between two S films of the same size $2L$. In this case the vector potential $A^\prime_m(x,h)$ in the upper superconducting electrode takes the form
\begin{equation}
A^\prime_m(x,h)=A_{0}-4Mh\left[{\rm arctg}\left(\frac{x+L}{h}\right)-{\rm arctg}\left(\frac{x-L}{h}\right)\right]. 
\end{equation}
This expression differs from Eq.~(9) by the additional constant $A_{0}$ chosen in a way that the total current $I_{tot}=\int_{-L}^L j_{sy}(x,h)dx$ flowing along the upper electrode is zero:
\begin{equation}\label{A0_res}
A_{0}=8Mh\left[{\rm arctg}\left(\frac{2L}{h}\right)-\frac{h}{2L}{\rm ln}\left(\frac{4L^2+h^2}{h^2}\right)\right]. 
\end{equation}
Then one may obtain the expression for the vector potential $A_s(x,z)$ similar to Eq.~\ref{IEAs} which takes the form
\begin{equation}\label{IEAsP}
A_s(x,z)=\int\limits_{-L}^L \frac{A^\prime_m(x^\prime,h)}{4\pi\lambda_{\rm eff}} {\rm ln}\left[\frac{(x-x^\prime)^2+(z-h)^2}{(x-x^\prime)^2+(z+h)^2}\right]dx^\prime. 
\end{equation}
Performing the integration we find that in the limit $L\gg h$ the $x$-component of the magnetic field ${\bf B}_s(x,0)$ in the cental part of the F slab (i.e. at $\left|x\right|\ll L$) reads
\begin{equation}\label{Bsx_res_zero}
B_{sx}(x,0)\approx -8M\left(\frac{h}{\lambda _{\mathrm{eff}}}\right)\left(\frac{h}{L}\right)\ln \left(\frac{2L}{h}\right).
\end{equation}

\subsection{Electromagnetic proximity effect in S/F and S/F/S systems}

For S/F bilayer shown in Fig.~3(a) the solution of the London equation for the magnetic field component $B_x(z)$ satisfying the boundary conditions has the form
\begin{equation}\label{supp_SF}
B_x=\left\{ 
\begin{array}{l}{\displaystyle 4\pi M \cosh\left(\frac{z}{\lambda_f}\right)+W_1\sinh\left(\frac{z}{\lambda_f}\right), ~0<z<h,}
\\{}\\{\displaystyle W_2\sinh\left(\frac{z-h-d_s}{\lambda_s}\right), ~h<z<h+d_s,}\end{array}
\right.
\end{equation}
where the constants $W_1$ and $W_2$ read
\begin{equation}\label{SF_W1} 
W_1=4\pi M\frac{\displaystyle 1-\cosh\left(\frac{h}{\lambda_f}\right)-\frac{\lambda_f}{\lambda_s}\sinh\left(\frac{h}{\lambda_f}\right)\tanh\left(\frac{d_s}{\lambda_s}\right)}{\displaystyle \sinh\left(\frac{h}{\lambda_f}\right)+\frac{\lambda_f}{\lambda_s}\cosh\left(\frac{h}{\lambda_f}\right)\tanh\left(\frac{d_s}{\lambda_s}\right)},
\end{equation} 
\begin{equation}\label{SF_W2} 
W_2=\frac{\displaystyle  4\pi M\left[\cosh\left(\frac{h}{\lambda_f}\right)-1\right]}{\displaystyle \frac{\lambda_s}{\lambda_f}\sinh\left(\frac{h}{\lambda_f}\right)\cosh\left(\frac{d_s}{\lambda_s}\right)+\cosh\left(\frac{h}{\lambda_f}\right)\sinh\left(\frac{d_s}{\lambda_s}\right)}.
\end{equation} 
In the limit $h\ll\lambda_f$ and $d_s\sim \lambda_s$ one gets $W_1\approx -4\pi M (h/\lambda_f)$ and, consequently, the magnetic field at the S/F interface (i. e. at $z=h$) is approximately equal to 
\begin{equation}\label{SF_B_res}
B_x(h)\approx 4\pi M \left[1-\frac{1}{2}\left(\frac{h}{\lambda_f}\right)^2\right].
\end{equation}
Remarkably, the demagnetization field is determined mainly by the London penetration depth $\lambda_f$ inside the ferromagnet.

For the layered S/F/S structure shown in Fig.~3(b) the solution of the London equation for $z>0$ gives   
\begin{equation}\label{supp_SFS}
B_x=\left\{ 
\begin{array}{l}{\displaystyle W_3  \cosh\left(\frac{z}{\lambda_f}\right), ~0<z<h,}
\\{}\\{\displaystyle W_4\sinh\left(\frac{z-h-d_s}{\lambda_s}\right), ~h<z<h+d_s,}\end{array}
\right.
\end{equation}
and $B_x(-z)=B_x(z)$. The constants $W_3$ and $W_4$ read
\begin{equation}\label{SF_W3} 
W_3=\frac{\displaystyle  4\pi M}{\displaystyle \cosh\left(\frac{h}{\lambda_f}\right)+\frac{\lambda_f}{\lambda_s}\sinh\left(\frac{h}{\lambda_f}\right)\tanh\left(\frac{d_s}{\lambda_s}\right)},
\end{equation} 
\begin{equation}\label{SF_W4} 
W_4=\frac{\displaystyle  4\pi M\sinh\left(\frac{h}{\lambda_f}\right)}{\displaystyle \frac{\lambda_s}{\lambda_f}\cosh\left(\frac{h}{\lambda_f}\right)\cosh\left(\frac{d_s}{\lambda_s}\right)+\sinh\left(\frac{h}{\lambda_f}\right)\sinh\left(\frac{d_s}{\lambda_s}\right)}.
\end{equation} 
In the limit $h\ll\lambda_f$ and $d_s\sim\lambda_s$ one gets 
\begin{equation}
\displaystyle W_3\approx \frac{4\pi M}{1+\left(h/\lambda_s\right)\tanh\left(d_s/\lambda_s\right)}.
\end{equation} 
Note that in contrast to the S/F structures where the demagnetization field is determined by $\lambda_f$ in S/F/S trilayers the screening of the magnetic field is defined by the London penetration depth $\lambda_s$ inside the superconductor. Then the magnetic field at the S/F interface (i. e. at $z=h$) is approximately equal to 
\begin{equation}\label{SFS_B_res}
B_x(h)\approx 4\pi M \left[1-\frac{h}{\lambda_s}\tanh\left(\frac{d_s}{\lambda_s}\right)\right].
\end{equation}

Comparing Eqs.~(\ref{SF_B_res}) and (\ref{SFS_B_res}) one sees that the screening in the S/F/S structures should be much larger than in S/F bilayers.

\end{document}